\documentclass[preprint,superscriptaddress,citeautoscript]{revtex4-1}
\usepackage{graphicx}
\setcitestyle{super}
\begin{document}

\title{Bright single-photon sources in bottom-up tailored nanowires}

%% Notice placement of commas and superscripts and use of &
%% in the author list

\newcommand{\TUDelft}{Kavli Institute of Nanoscience, Delft University of Technology, Delft, The Netherlands}
\newcommand{\Eindhoven}{Eindhoven University of Technology, Eindhoven, The Netherlands}
\newcommand{\jointauthors}{These authors contributed equally to this work}
\renewcommand{\abstractname}{}

\author{Michael E. Reimer$^\dagger$}
    \email{m.e.reimer@tudelft.nl}
    \affiliation{\TUDelft}
    \altaffiliation{\jointauthors}
\author{Gabriele Bulgarini$^\dagger$}
    \affiliation{\TUDelft}
\author{Nika Akopian}
    \affiliation{\TUDelft}
\author{Mo\"{i}ra Hocevar}
    \affiliation{\TUDelft}
\author{Maaike~Bouwes~Bavinck}
    \affiliation{\TUDelft}
\author{Marcel~A.~Verheijen}
    \affiliation{\Eindhoven}
\author{Erik~P.~A.~M.~Bakkers}
   \affiliation{\TUDelft}
   \affiliation{\Eindhoven}
\author{Leo~P.~Kouwenhoven}
   \affiliation{\TUDelft}
\author{Val~Zwiller}
   \affiliation{\TUDelft}

\date{\today}

\begin{abstract}
\textbf{The ability to achieve near-unity light extraction efficiency is necessary for a truly deterministic single photon source. The most promising method to reach such high efficiencies is based on embedding single photon emitters in tapered photonic waveguides defined by top-down etching techniques. However, light extraction efficiencies in current top-down approaches are limited by fabrication imperfections and etching induced defects. The efficiency is further tempered by randomly positioned off-axis quantum emitters. Here, we present perfectly positioned single quantum dots on the axis of a tailored nanowire waveguide using bottom-up growth. In comparison to quantum dots in nanowires without waveguide, we demonstrate a 24-fold enhancement in the single photon flux, corresponding to a light extraction efficiency of 42\,$\%$. Such high efficiencies in one-dimensional nanowires are promising to transfer quantum information over large distances between remote stationary qubits using flying qubits within the same nanowire p-n junction.
}
\end{abstract}

\maketitle

The ability to achieve near-unity light extraction efficiency is necessary for a truly deterministic single photon source and would revolutionize the fields of quantum computation\cite{Kni01}, quantum cryptography\cite{Gisin02} and quantum optics\cite{Strauf10}. Recently, significant progress has been made toward achieving high efficiency single photon sources\cite{Pelton02,Strauf07,Claudon10,Dal10,Bab10,Lee11}. Amongst these sources, semiconductor quantum dots show particular promise since they can be embedded in devices enabling electrical injection, while maintaining efficient light emission \cite{Heindel10, Gregersen10}. The most frequent approach to enhance the light extraction efficiency $\eta$ of a single photon source, defined as the fraction of photons collected by the first lens, is to embed the quantum dots in microcavities and utilize the Purcell effect when the exciton emission is resonant with a cavity mode. However, to reach a high efficiency, a high quality factor and low mode volume is required, which therefore limits the bandwidth to a fraction of a nanometer in wavelength\cite{Strauf10}. Moreover, $\eta$ is further limited due to losses such as fabrication imperfections or absorption losses in the cavity\cite{Barnes02}.

A more recent approach that overcomes the narrow bandwidth limitation of microcavities and thus allows for broadband spontaneous emission control is based on tapered photonic waveguides \cite{Claudon10}. In that work, the tapered photonic waveguides were achieved by etching a GaAs wafer containing multiple InAs quantum dots. The light extraction efficiency was increased by optimizing the photonic waveguide diameter to maximize the amount of light emitted from the quantum dot into the fundamental mode of the waveguide as compared to higher-order modes, and by tapering the waveguide tip to minimize total internal reflection at the semiconductor-air interface\cite{Claudon10}. Such reflections were avoided due to a smooth reduction in the effective refractive index towards the nanowire tip and adiabatic expansion of the fundamental mode during propagation. An additional advantage of the nanowire taper is the good directionality of photon emission from the nanowire, which allows for efficient collection by a commercial objective\cite{Claudon10}. However, due to the random positions of the quantum dots, there are multiple dots in each individual etched waveguide. Consequently, the quantum dots are not likely to be found exactly on the waveguide axis, which diminishes the coupling of the emitter to the fundamental waveguide mode, thus drastically reducing the overall light extraction efficiency \cite{Bleuse11}.

In this work, we increase the light extraction efficiency of a quantum dot emitter in a tailored nanowire waveguide using a bottom-up growth approach through independent control of both the nanowire shape and quantum dot location. Under the appropriate growth conditions, we have positioned a single InAsP quantum dot exactly on the axis of an InP nanowire waveguide with a very small tapering angle towards the tip (${\alpha=2^\circ}$). Our bottom-up approach has the potential to reach near-unity light extraction efficiency ($\eta=97\%$), assuming a perfect mirror below the nanowire, a ${1^\circ}$ nanowire taper and high coupling of the quantum emitter into the fundamental waveguide mode compared to higher order modes (i.e., $\beta$-factor of 0.95) for a quantum dot emitter on-axis of the nanowire waveguide \cite{Friedler09}. Single, on-axis quantum emitters and reduced tapering angles are possible with our nanowire tailoring bottom-up growth technique and out of reach of current top-down etching techniques \cite{Greg08,Claudon10}.

\section*{Results}
\subsection*{Tailored nanowire waveguide geometry using bottom-up growth.}
\begin{figure}[h]
  \centering
  % Requires \usepackage{graphicx}
  \includegraphics[width=8.25cm]{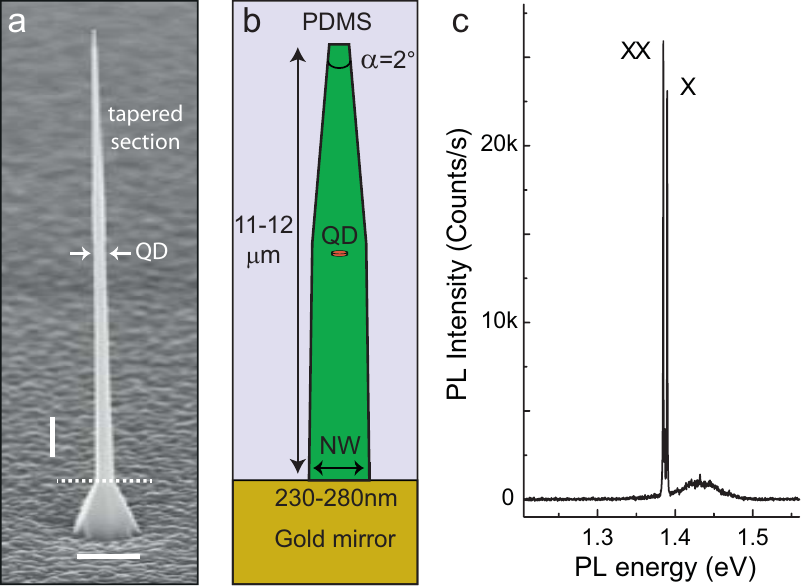}
  \caption{\textbf{Tailored nanowire waveguide geometry.} (a) SEM image of tapered InP nanowire waveguide containing a single InAsP quantum dot (QD). The tapering angle of the nanowire is $\alpha=2^{\circ}.$ The nanowire breaks at the dotted line after transfer into PDMS. Vertical and horizontal scale bar: 1\,$\mu$m. (b) Tailored nanowire geometry embedded in PDMS with bottom gold mirror. The quantum dot (QD) is indicated by the red dot and, by design, is positioned on the nanowire waveguide axis where the emission is efficiently coupled to the fundamental waveguide mode. (c) Typical single dot PL of exciton ($X$) and biexciton ($XX$) in an individual tapered nanowire waveguide, exhibiting a clean optical spectrum over a large energy range. The broad peak at 1.43\,eV originates from the InP nanowire (NW).}\label{Fig:1}
\end{figure}

Fig. \ref{Fig:1}(a) shows an SEM image of our tailored nanowire waveguide with nanowire diameter of 200\,nm, length of 11.2\,$\mu m$ and 2\,$^{\circ}$ taper at the tip. Each individual nanowire contains, by design, a single quantum dot exactly on the nanowire axis. To achieve such a high level of control, we first grow the nanowire core, which consists of a thin segment of InAsP (quantum dot) embedded in an InP nanowire. At this stage, the nanowire and quantum dot have a diameter that is approximately equal to the size of the gold particle ($\sim20$\,nm) used as a catalyst in the vapour-liquid-solid growth mechanism. After growth of the quantum dot and nanowire, the temperature is increased in order to suppress axial growth and favor shell growth. The final result is a waveguide section at the quantum dot position and a very small nanowire taper towards the tip (see Methods). These two effects combined together are used to efficiently extract the light of our single photon emitter.

Since single photons emitted by the quantum dot can travel both directions in the nanowire waveguide, losses can be reduced by adding a metallic mirror at the nanowire base to reflect downward emitted photons. As our growth technique does not allow for growth of nanowires on top of a metallic mirror, we have integrated a gold mirror at the nanowire base by transferring them into a flexible and fully transparent polydimethylsiloxane (PDMS) polymer film as sketched in Fig. \ref{Fig:1}(b). The first section of the nanowire before quantum dot growth exhibits a small tapering angle of 0.4$^{\circ}$, whereas the second section of the nanowire after quantum dot growth features a tapering angle of $\alpha=1.5^{\circ}\pm$0.2$^{\circ}$. From SEM analysis of the as-grown nanowire sample, the nanowire diameter that surrounds the quantum dot was determined to be 160--220\,nm. We conclude from finite-difference time-domain (FDTD) simulations of InP nanowires, to be presented later, that the most efficient coupling of quantum dot emission into the fundamental mode in the nanowire waveguide for emission wavelength, $\lambda$ of 950\,nm occurs for a nanowire diameter, $D$ of 220\,nm (i.e., $D/\lambda=0.23$).

A photoluminescence (PL) spectrum from a single InAsP quantum dot in an InP tapered nanowire is presented in Fig. \ref{Fig:1}(c). Such behaviour illustrates a significant advantage of quantum dots in nanowires \cite{Borg05,vanWeert09,Heinrich10,Dal11} as compared to randomly grown quantum dots by the self-assembly growth process \cite{Pelton02,Strauf07,Claudon10}. Since the nanowire contains only a single quantum dot by nature of the bottom-up growth process, the observed PL spectrum exhibits only emission from a single quantum dot over a large energy range, originating from exciton and biexciton recombination. The broad emission peak at higher energy originates from the InP nanowire. Most important is the natural location of the quantum dot on the nanowire axis, which therefore couples to the optical field of a waveguide more efficiently than randomly positioned quantum dots \cite{Claudon10, Bleuse11}.

We verify that the quantum dot is exactly positioned on the nanowire waveguide axis by high resolution scanning transmission electron microscopy (STEM) images, as well as analysis of the chemical composition, presented in Fig. \ref{Fig:TEM}(a-c). As it is very difficult to image the quantum dot when surrounded by a thick shell of approximately 200\,nm, a nanowire sample with a thinner shell of approximately 100\,nm is measured. The STEM image in Fig. \ref{Fig:TEM}(a) is used to exhibit the thickness of the core and shell. Fig. \ref{Fig:TEM}(b) shows the same STEM image as in Fig. \ref{Fig:TEM}(a), but now using contrast/brightness settings optimized for visibility of the quantum dot. The quantum dot is of the pure wurtzite structure without stacking faults. From the resulting image, a quantum dot height of $7\pm1$\,nm and diameter of 20\,nm is obtained.

Confirmation of the quantum dot location is established by performing an energy-dispersive X-ray (EDX) spectroscopy line scan along the nanowire growth axis, indicated by the arrow in Fig. \ref{Fig:TEM}(b). The results of the EDX line scan are shown in Fig. \ref{Fig:TEM}(c), which shows an increase in the As concentration at the location of the quantum dot we infer from Fig. \ref{Fig:TEM}(b). By subtracting the contribution of the shell around the quantum dot, we conclude that the As concentration of the quantum dot is 12.5\,atomic\,$\%$, corresponding to a chemical composition of InAs$_{0.25}$P$_{75}$ (see Methods).

\subsection*{Single photon collection efficiency.}
\begin{figure}[h]
  \centering
  % Requires \usepackage{graphicx}
  \includegraphics[width=8.25cm]{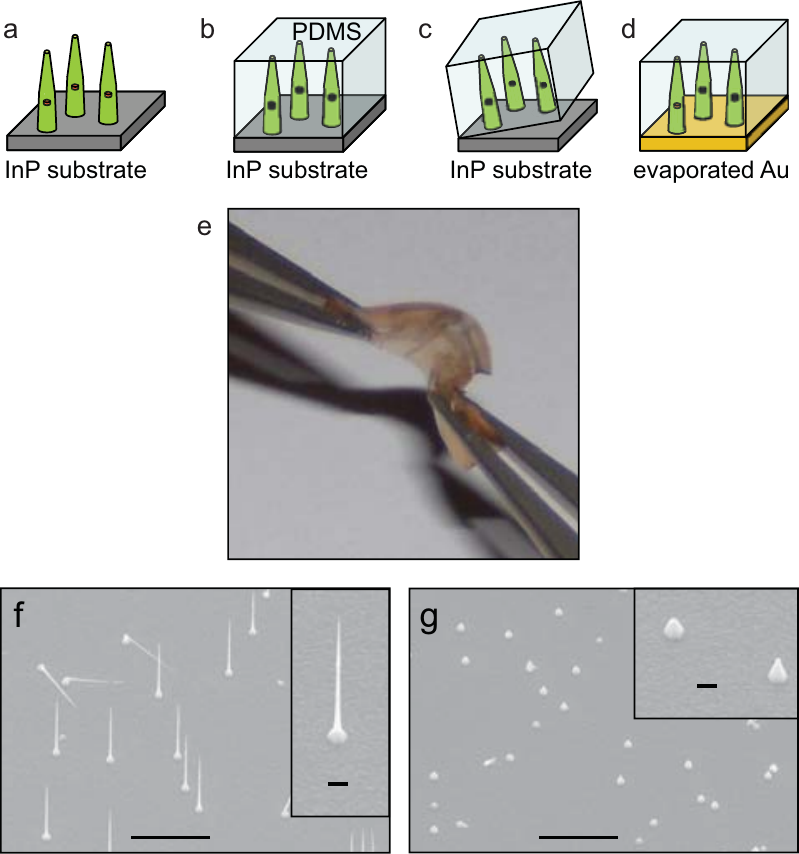}
  \caption{\textbf{PDMS nanowire transfer method for integration of bottom gold mirror.} (a) Growth of single InAsP quantum dots embedded in InP nanowires on an InP substrate. (b) PDMS is spin-coated and subsequently cured to obtain a thin layer ($\sim$10\,$\mu$m) comparable to the nanowire length. (c) Embedded nanowires in PDMS are removed from the InP substrate using tweezers and a razor blade to scrape the substrate surface. (d) An 80\,nm gold layer is evaporated at the nanowire base. The red dots in (a)-(d) indicate the quantum dots in individual nanowires. (e) Optical image of finished flexible device containing very bright single photon emitters, which are held between two tweezers. (f,g) SEM images of the as-grown sample before (f) and after (g) nanowire transfer. Since there are no nanowires remaining in (g), we conclude that 100$\%$ of nanowires are successfully transferred into PDMS. Scale bar: 10\,$\mu$m. The insets of (f) and (g) show a magnified view of their respective image with scale bar of 1\,$\mu$m.}\label{Fig:2}
\end{figure}

A gold mirror is integrated at the nanowire base by transferring the nanowires into PDMS, followed by metal evaporation (see Methods and Fig. \ref{Fig:2}(a)-(e) for details). An SEM image of the as-grown sample is shown before (Fig. \ref{Fig:2}(f)) and after (Fig. \ref{Fig:2}(g)) the nanowire transfer process. Since only the thick nanowire bases are left in Fig. \ref{Fig:2}(g), we conclude that 100 $\%$ of nanowires are transferred successfully into PDMS. The wires in Fig. \ref{Fig:2}(f) that appear to be lying on the substrate are a small fraction of wires that are kinked and continue to grow along other crystallographic directions instead of the normal (111).

\begin{figure}[h]
  \centering
  % Requires \usepackage{graphicx}
 \includegraphics[width=8.25cm]{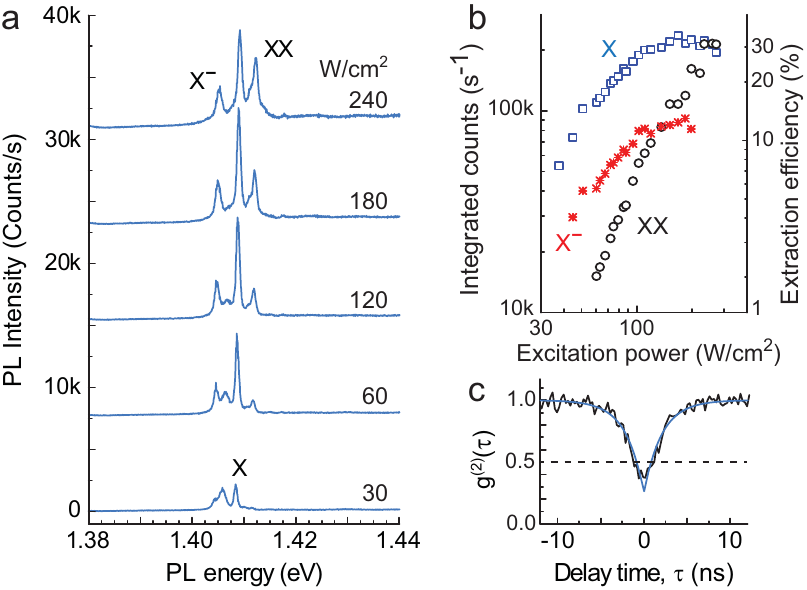}
  \caption{\textbf{Bright single photon emission of tailored nanowire waveguides.} (a) Typical power dependence of single quantum dot in tailored nanowire waveguide under pulsed laser excitation. The PL spectra are offset for clarity. (b) Integrated counts of exciton ($X$), biexciton ($XX$), and charged exciton ($X^-$) from (a) as a function of increasing excitation power. Remarkably, the $X$ line saturates at values as high as 236,000 counts per second measured on the CCD detector, resulting in a photon collection efficiency of $\sim30\%$. Accounting for competition between recombination of $X^-$ and $X$, we estimate an $\eta$ for the nanowire waveguide of 42\,$\%$. We note that the calculated collection efficiencies are only valid at quantum dot saturation where the generation efficiency is maximum. (c) Second order correlation function, g$^{(2)}(\tau$), of single $X$ line. The anti-bunching dip below 0.5 is characteristic of a single photon emitter. Accounting for the finite time response of our single photon detectors, we obtain $g^{(2)}(0)$ of 0.12 and $X$ lifetime of 1.7\,ns$\pm$0.1\,ns.}\label{Fig:3}
\end{figure}

We characterize the collection efficiency enhancement of our single photon emitter through PL studies under pulsed laser excitation at a repetition rate of 76\,MHz. Pulsed excitation provides a good estimate of the extraction efficiency since it is expected that the quantum dot returns to the ground state and cannot emit until the next excitation pulse. Power dependent PL spectra for a single quantum dot after transfer into PDMS and integration of the bottom gold mirror are shown in Fig. \ref{Fig:3}(a). At low excitation power, the main PL peak observed at 1.409\,eV is attributed to the neutral exciton ($X$) involving electron-hole recombination in the quantum dot s-shell. Further increase of the excitation power results in biexciton ($XX$) emission, 3\,meV above the exciton line. In our nanowire samples, the biexciton emission can appear either above\cite{vanWeert09} or below\cite{Kouwen10,Reimer11} depending on the dot height. Both exciton and biexciton are identified by the observed linear and quadratic power dependence, respectively, as shown in Fig. \ref{Fig:3}(b). The identification of exciton and biexciton was further verified in other quantum dots on the same sample that exhibit a fine-structure splitting. Typical fine-structure splittings of the order of 30\,$\mu$eV are observed for InAsP quantum dots in nanowires \cite{vanWeert09}. The singly charged exciton, $X^-$, is 4\,meV below the exciton line and was determined from the observed linear power dependence in Fig. \ref{Fig:3}(b) and electric field measurements of dots in nanowires studied in earlier work. \cite{Kouwen10,Reimer11}

The brightness of our single photon source is presented in Fig. \ref{Fig:3}(b), which shows the integrated count rate from Fig. \ref{Fig:3}(a) as a function of excitation power. At saturation, the integrated count rate of the single exciton line measured on the CCD detector is 236,000 counts per second. The measured count rate is a factor of 24 times brighter than previously reported on InAsP quantum dots in InP nanowires \cite{vanWeert09}. Taking into account the measured collection efficiency of our experimental setup ($\sim1$\,$\%$), a single photon flux at the first lens of 24 MHz is estimated. This estimation results in a light extraction efficiency, $\eta$ of our single photon emitter of $\sim30$\,$\%$, corresponding to 0.3 photons per pulse arriving at the first lens. We note that since there is a competition between the charged exciton and neutral exciton states in the observed PL emission spectra, we in fact underestimate the efficiency of the nanowire waveguide. From the integrated intensity at  saturation (Fig. \ref{Fig:3}(b)), we estimate that approximately one-quarter of the time an electron is captured into the quantum dot resulting in recombination of the charged exciton instead of the neutral exciton. Taking into account the competition between recombination of charged exciton and neutral exciton states, our estimated collection efficiency of the nanowire waveguide is 42\,$\%$. The deviation from the expected 71$\%$ when assuming a 2\,$^{\circ}$ nanowire taper may be due to several factors such as the quality of the gold mirror, internal quantum efficiency of the quantum dot, and unwanted reflections at the PDMS-air interface. The main source of discrepancy between the theoretical and measured $\eta$ is due to the quality of the gold mirror. If we assume a modal reflectivity for the guided mode on the mirror of 30\,$\%$ with $D/\lambda=0.22$ according to calculations by Friedler et al.\cite{Friedler08}, and 4$\%$ total internal reflection losses at the PDMS-air interface, the expected $\eta$ matches the measured 42\,$\%$ for a quantum dot internal quantum efficiency of 95\,$\%$. In order to approach the theoretical $\eta$ of 71\,$\%$, the modal reflectivity of the guided mode on the mirror may be improved by adding a thin dielectric between the mirror and semiconductor nanowire as suggested by Friedler et al.\cite{Friedler08}, while losses at the PDMS-air interface can be diminished by using silica aerogel with an index of refraction close to 1.

In Fig. \ref{Fig:3}(c), we demonstrate single photon emission by filtering the single $X$ line close to saturation and perform photon correlation measurements in a standard Hanbury-Brown and Twiss setup. The second order correlation function $g^{(2)}(\tau)$ indicates strong multi-photon suppression evident by the absence of coincidence counts at zero time delay and is a signature of single photon emission. We measure $g^{(2)}(0)$ of 0.36. Taking into account the finite time response of the Si avalanche photodiodes, we obtain $g^{(2)}(0)$ of 0.12. The remaining contribution to $g^{(2)}(0)$ is due to background photons that originates from the tail of the InP nanowire emission. We have thus subtracted this background emission from our measured photon fluxes to estimate the collection efficiencies.

\subsection*{Modeling of fundamental mode in nanowire waveguide.}
\begin{figure}[h]
  \centering
  % Requires \usepackage{graphicx}
  \includegraphics[width=8.25cm]{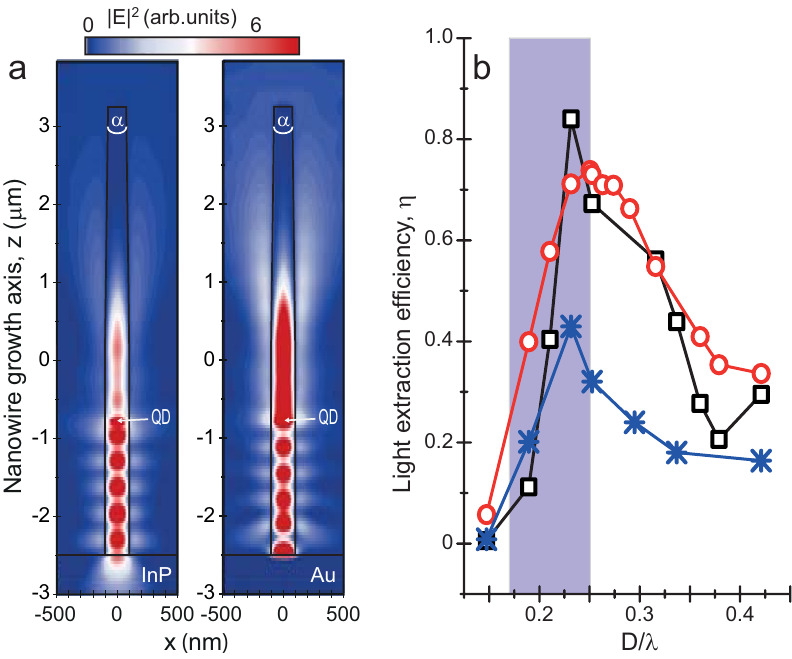}
  \caption{\textbf{Calculated light extraction efficiency from single quantum dot in tailored nanowire waveguides.} (a) Finite-difference time-domain simulations showing the intensity profile for a 2D-cut along the nanowire growth axis. The integrated intensity at the nanowire end is a factor of 2.7 larger on the gold (Au) substrate (right) as compared to the InP substrate (left). The nanowire dimensions are similar to the experimentally investigated nanowires using a tapered nanowire geometry with full-angle of $\alpha=2^{\circ}$ and nanowire diameter of 200\,nm. An in-plane electric dipole emitter is used to represent the quantum dot (QD). (b) Calculated light extraction efficiency, $\eta$, at nanowire end using an NA of 1 for tapered nanowire surrounded by air (black squares) or embedded in PDMS (red circles), both on top of a gold substrate. The absence of the nanowire taper (blue stars) reduces the maximum $\eta$ through reflections at the nanowire top. The shaded region indicates the range of $D/\lambda$ used in the present experiments.}\label{Fig:4}
\end{figure}

Finite-difference time-domain (FDTD) simulations were performed using a nanowire waveguide geometry similar to the experimentally investigated nanowires in order to quantify the light intensity enhancement due to the presence of bottom gold mirror that replaces the InP substrate during the nanowire transfer process. The results of these calculations, excluding PDMS, are shown in Fig. \ref{Fig:4}(a) for a nanowire on an InP and gold substrate. In both cases, the intensity profile is plotted for a two-dimensional cut along the nanowire growth axis for an in-plane electric dipole emitter, which is used to represent a flat quantum dot in III-V nanowires \cite{Niquet08}. In each case, a standing wave of the fundamental mode HE$_{11}$ is observed between the emitter and substrate for a nanowire diameter of 200\,nm and emission wavelength $\lambda$ of 950\,nm. The emitter location has been optimized by positioning it at one of the antinodes of the electric field. It is very important that the reflected light from the downward emitted photons is in-phase at the emitter position so that they interfere constructively. The light intensity at the nanowire end is determined from the simulations for a numerical aperture of 1 and increases by a factor of 2.7 with the addition of a gold substrate as compared to InP only. The increase in light emission at the nanowire end with gold substrate is evident of Fig. \ref{Fig:4}(a). Including PDMS in Fig. \ref{Fig:4}(b), the light enhancement of 2.7 is reduced to 2.4 as the effective index of refraction surrounding the electric dipole is modified. The deviation from the intuitive factor of 2 is because the presence of the gold mirror modifies the electromagnetic environment of the quantum emitter. As a result, the single photon flux is increased due to a shortening of the quantum dot exciton radiative lifetime for a quantum dot positioned at the electric field anti-node \cite{Bleuse11}.

In the following, we optimize the calculated light extraction efficiency, $\eta$, at the nanowire end for a numerical aperture (NA) of 1 by modifying the nanowire diameter with (red circles) and without (black squares) PDMS on top of a gold substrate (Fig. \ref{Fig:4}(b)). The optimum $\eta$ only shifts from $D/\lambda$ of 0.23 without PDMS to $D/\lambda$ of 0.25 with addition of PDMS as the effective refractive index of PDMS (n=1.5) is slightly larger than air. These $D/\lambda$ values are within the range of nanowires investigated in this work, indicated by the shaded region in Fig. \ref{Fig:4}(b). For comparison, we also show $\eta$ without nanowire tapering in air (blue stars). The effect of the tapered nanowire tip results in a $\sim2$-fold enhancement in the maximum $\eta$ as compared to no nanowire taper. By varying the NA in the FDTD simulations from 1 to 0.75 (NA of our objective in experiment), there is only a reduction of $\sim3$\,$\%$ in the calculated $\eta$ due to the good directionality of photon emission from the nanowire taper.

\subsection*{Collection efficiency enhancement.}
\begin{figure}[h]
  \centering
  % Requires \usepackage{graphicx}
  \includegraphics[width=8.25cm]{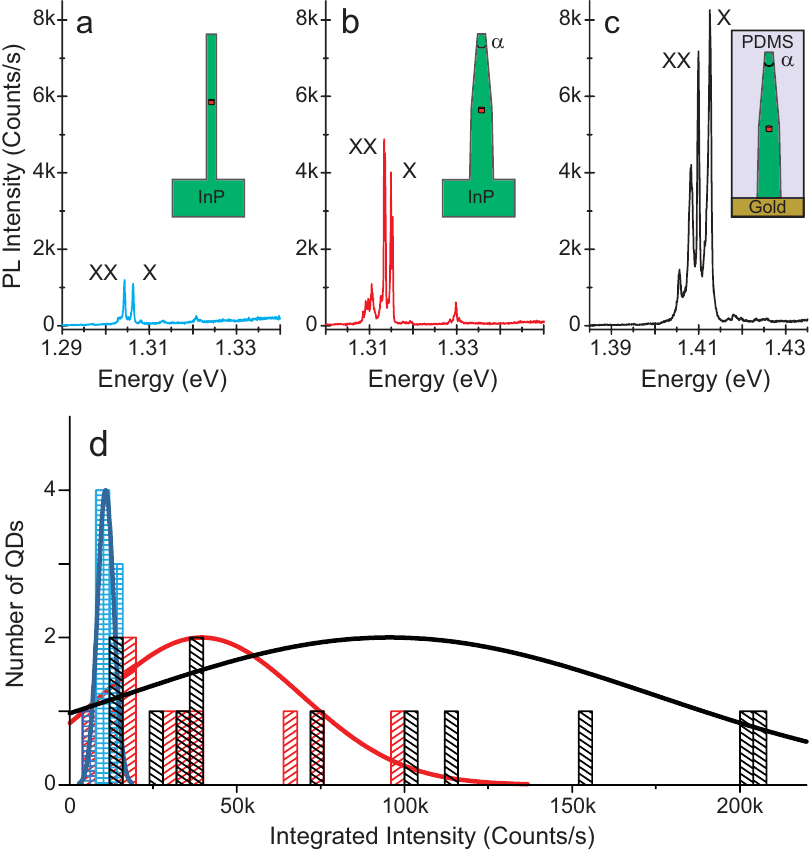}
  \caption{\textbf{Enhanced single quantum dot emission.} Comparison of PL intensity for the brightest quantum dots obtained utilizing three different nanowire geometries: (a) no waveguide; (b) tapered nanowire waveguide; and (c) tapered nanowire waveguide + gold + PDMS. The integrated PL intensity of the exciton ($X$) increases by a factor of up to 20 with the tapered nanowire waveguide and bottom gold mirror in (c) as compared to no waveguide in (a). The inset of (a)-(c) depicts schematically the nanowire geometry. $XX$: biexciton. (d) Histogram of integrated PL Intensity of single exciton ($X$) peak at saturation. The y-axis represents the number of quantum dots with measured PL intensity for a bin size of 4,000 counts/s. The dispersion of data originates from the variable distance of the quantum dot to the mirror. The solid lines are Gaussian fits to the probability distribution and serve as an illustration of the increased trends by our technique. Blue: no waveguide; red: tapered nanowire waveguide; black: tapered nanowire waveguide + gold + PDMS. Statistically, there is an observed enhancement of 2.4 with tapered nanowire waveguide and gold mirror in comparison to no mirror, which is in good agreement with our simulations.}\label{Fig:5}
\end{figure}

In Fig. \ref{Fig:5}, we demonstrate an enhancement of the single quantum dot PL intensity with our tailored nanowires in comparison to non-tailored nanowires. The comparison is made for our brightest quantum dot obtained for three different nanowire geometries: no waveguide (Fig. \ref{Fig:5}(a)); tapered nanowire waveguide (Fig. \ref{Fig:5}(b)); and tapered nanowire waveguide with gold mirror and PDMS (Fig. \ref{Fig:5}(c)). The single exciton ($X$) integrated PL intensity exhibits a clear enhancement with optimum tapered nanowire waveguide and further increase with integration of gold mirror in comparison to no waveguide. The main factors that influence the observed enhancement are nanowire diameter and quantum dot position along the nanowire growth axis. When these conditions are fulfilled, the single photon emission flux is enhanced 20-fold with tapered nanowire waveguide and gold mirror (Fig. \ref{Fig:5}(c)) as compared to no waveguide (Fig. \ref{Fig:5}(a)). Since we cannot measure the exact same quantum dot for all three cases, a statistical distribution of integrated PL intensity for the single exciton ($X$) line at saturation for ten quantum dots is shown in Fig. \ref{Fig:5}(d). A clear trend in the observed distribution of integrated PL intensity is visible; however, we observe a broader distribution using the waveguide and gold mirror as compared to no waveguide. The broader distribution is due to the variation in the nanowire diameter and distance of the quantum dot emitter from the mirror. From the maximum of the probability distribution fit in Fig. \ref{Fig:5}(d), we obtain a factor of 2.4 increase in single photon emission from the tapered nanowire waveguide and gold mirror as compared to waveguide only. The average enhancement of 2.4 with the gold mirror and waveguide compared to waveguide only is in excellent agreement with the theoretical predictions from Fig. \ref{Fig:4}.

\section*{Discussion}
The 20-fold enhancement in single photon emission that we observe for individual quantum dots in tailored nanowire geometries using a bottom-up growth approach compared to non-tailored nanowire geometries makes future experiments in the combination of both transport and optics more realistic. As an example, a single quantum dot embedded in a nanowire waveguide can be coupled to an electrostatic defined quantum dot\cite{Nadj10} within the same nanowire to coherently convert a single electron spin into the polarization state of a photon. Coherent spin to photon conversion makes the long-distance transfer of quantum information between remote stationary qubits a possibility. Additionally, the nanowire transfer technique presented here using PDMS is promising for the realization of efficient electrically driven single photon sources \cite{Minot07} and solar cells \cite{Kel10} based on one-dimensional nanowires with controlled shape and tapering.

Finally, we comment on possible improvements to our tailored nanowire geometry in order to achieve light extraction efficiencies exceeding 90\,$\%$ in future work. First, the quality of the gold mirror may be improved by adding a thin dielectric layer between the semiconductor and nanowire as suggested by Friedler et al.\cite{Friedler08}. Second, the quantum dot blinking from neutral to charged exciton states can be circumvented by precise control of the charge state with the use of electrostatic gates positioned around an individual nanowire \cite{Kouwen10,Reimer11} such that a well-defined frequency of the single photon source is attained. Lastly, the distance between the quantum dot and cleaved nanowire interface, and thus the gold mirror should be precisely controlled to ensure that the quantum dot is positioned at the anti-node of the electric field. Combining these suggested improvements with the fact that our quantum dot is always on the nanowire waveguide axis, indeed, anticipates very high count rates to be achieved.

\textbf{Methods:}\\
\textbf{Nanowire growth.} InP nanowires were grown for 40\,min at a temperature of 420\,$^{\circ}$C by the vapour-liquid-solid (VLS) growth mechanism in a low pressure metal-organic vapor-phase epitaxy (MOVPE) reactor. The VLS mechanism requires a metal catalyst particle. In our case, 20 nm gold colloids were dispersed on a (111)B InP substrate. The precursors used for InP growth were trimethyl-indium and phosphine. After 20\,min of growth, an InAs$_{25}$P$_{75}$ quantum dot was incorporated by addition of As in the MOVPE reactor using an arsine flux. The quantum dot height was set by the growth time (1-2 seconds), and is about 7\,nm. After nanowire growth, the temperature was raised up to 500\,$^{\circ}$C to favor radial versus axial growth. The shell growth time and temperature were carefully chosen to shape the nanowire geometry and obtain an optimum nanowire diameter and tapering angle towards the tip. The total length of the nanowire is typically 11--12\,$\mu$m, and the nanowire diameters at the quantum dot position and the tip are 160--220\,nm and 20--30\,nm, respectively. As a result, the nanowire waveguiding section at the quantum dot position has $D/\lambda$ values ranging from 0.17 to 0.25 and the nanowire tapering angle at the tip is $\alpha=1.5^{\circ}\pm0.2^{\circ}$.

\begin{figure}[h]
  \centering
  % Requires \usepackage{graphicx}
  \includegraphics[width=12.5cm]{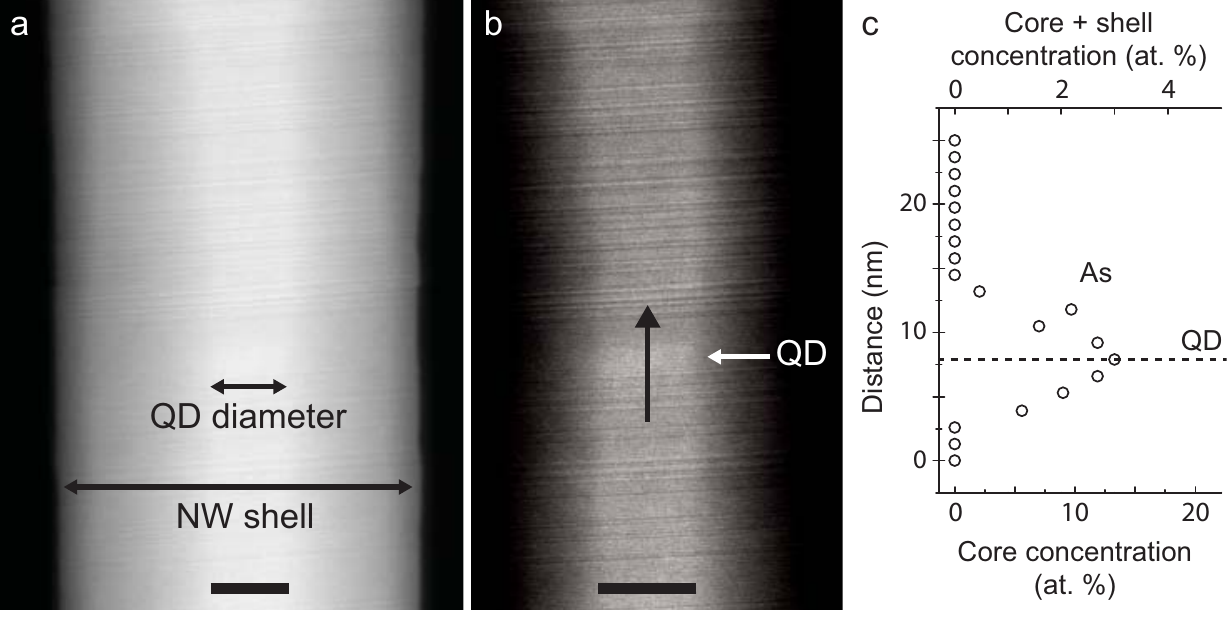}
  \caption{\textbf{Single quantum dot on the nanowire waveguide axis.} High angle angular dark field (HAADF) scanning transmission electron microscopy images confirming that the quantum dot is perfectly positioned on the nanowire waveguide axis. Optimized contrast settings for visibility of nanowire (NW) core and shell in (a) and quantum dot in (b). The quantum dot is recognized along the black arrow in (b) by the contrast change from dark to bright regions along the nanowire. Scale bar: 20\,nm. (c) Energy-dispersive X-ray spectroscopy line scan along the arrow in (b) confirming the presence of the quantum dot. The As concentration reaches a maximum of 12.5\,$\%$ in the core only, thus resulting in a chemical composition for the quantum dot of InAs$_{0.25}$P$_{0.75}$.}\label{Fig:TEM}
\end{figure}

\textbf{Quantum dot composition.} The As concentration of the quantum dot is confirmed by independent measurements based on the resulting intensities obtained from the high angle angular dark field (HAADF) STEM images presented in Fig. \ref{Fig:TEM}. At the position of the quantum dot, the brightness after background subtraction is 2.8\,$\%$ higher. Since the HAADF intensity scales with $tZ^{}$, where $t$ is the local sample thickness and $Z$ is the atomic number, thus, the composition can be determined. The total brightness consists of both the core and shell contributions. Using the HAADF intensity, $I =tZ_{average}^{1.7}$, where $Z_{average} = \{Z_{In}+xZ_{As}+(1-x)Z_P\}/3$, we find that $x = 0.26 \pm 0.06$, which is in excellent agreement with the quantum dot composition of InAs$_{25}$P$_{75}$ obtained from the EDX results.

\textbf{Single quantum dot PL.} Individual nanowires were isolated for single quantum dot PL by using a low density colloid solution ($\sim0.3$\,$\mu$m$^2$).  The single quantum dot PL was measured in a standard micro-photoluminescence (PL) setup using non-resonant continuous excitation at wavelength of 532\,nm and laser spot size of $\sim$ 1$\mu$m. In the measurements to characterize the light extraction efficiency, a 3\,ps pulsed Ti:sapphire laser with repetition rate of 76\,MHz and wavelength of 744\,nm was utilized. The nanowires containing quantum dots are cooled down to 10\,K in a continuous flow cryostat and the emitted photons are collected by a microscope objective with NA of 0.75. A flip mirror is used to direct the emitted photons either to a Hanbury-Brown and Twiss setup for photon correlation measurements or to a spectrometer with a nitrogen cooled Si CCD detector for spectral analysis.

\textbf{PDMS film.}
The preparation of the polydimethylsiloxane (PDMS) polymer solution consists of mixing a polymer base and a curing agent with a weight ratio of 10:1. The polymer used is Sylgard 184 PDMS elastomer kit from Dow Corning. The uncured polymer was diluted with methylene chloride (0.5\,g/ml) in order to decrease the viscosity of the polymer and enable the realization of thin films by spin coating the polymer solution on the nanowire substrate \cite{Plass09}. Using a spin coating speed of 6000 rpm for 1 minute, a 10\,$\mu$m layer of PDMS was obtained. This thickness of the PDMS film is comparable to the length of the nanowires. After depositing the PDMS film, a waiting time of 15 hours was used to allow the polymer to set. This process improved the peeling of the polymer film containing the nanowires after curing at 125\,$^{\circ}$C for 20 minutes.

%\bibliography{PDMSbib}

%%%%%%%%%%%%%%%%%%%% ACKNOWLEDGMENTS %%%%%%%%%%%%%%%%%%%%%%%%%%%%%%%%

\begin{acknowledgments}
We acknowledge Andika Asyuda for contribution to the development of the nanowire transfer technique into the PDMS polymer film and Reinier Heeres for useful discussions pertaining to the FDTD simulations.  This work was supported by the Netherlands Organization for Scientific Research (NWO), Dutch Organization for Fundamental Research on Matter (FOM), European Research Council and DARPA QUEST grant.
\end{acknowledgments}

\noindent
\textbf{Author contributions}
The experiments were conceived and designed by M.E.R and V.Z. and carried out by M.E.R., G.B. and N.A. The data were analyzed by M.E.R. and G.B. The simulations were carried out by M.B.B. The sample was developed by M.H. and E.P.A.M.B. The manuscript was written by M.E.R. with input from G.B., N.A., M.H., M.B.B., M.A.V., E.P.A.M.B., L.P.K. and V.Z.

\noindent
\textbf{Competing financial interests:} The authors declare no competing financial interests.

\newpage
%%%%%%%%%%%%%%%%
%Figure 1
%%%%%%%%%%%%%%%%

%%%%%%%%%%%%%%%%
%Single dot on waveguide axis
%%%%%%%%%%%%%%%%

%%%%%%%%%%%%%%%%
%Figure 2
%%%%%%%%%%%%%%%%

%%%%%%%%%%%%%%%%
%Figure 3
%%%%%%%%%%%%%%%%

%%%%%%%%%%%%%%%%
%Figure 4
%%%%%%%%%%%%%%%%
%%%%%%%%%%%%%%%%
%Figure 5
%%%%%%%%%%%%%%%%

%% Put the bibliography here, most people will use BiBTeX in
%% which case the environment below should be replaced with
%% the \bibliography{} command.
%\begin{thebibliography}{1}
%\bibitem{dummy} Articles are restricted to 50 references, Letters
%to 30.
%\bibitem{dummyb} No compound references -- only one source per
%reference.
%\end{thebibliography}

\end{document}